# Applications of DMDs for Astrophysical Research

M. Robberto*[a], A. Cimatti[b], A. Jacobsen[c], F. Zamkotsian[d], F. M. Zerbi[e]

[a]Space Telescope Science Institute, 3700 San Martin Dr., Baltimore, MD, USA 21218;
[b]Dipartimento di Astronomia, Università di Bologna, Via Ranzani 1, I-40127, Bologna, Italy
[c]OpSys Project Consulting, Hauptstrasse 3A, D-35641 Schoeffengrund, Germany
[d]Lab. d'Astrophysique de Marseille, 38 rue F. Joliot Curie, 13388 Marseille Cedex 13, France
[d]INAF Osservatorio Astronomico di Brera, Via Bianchi 46, I-23807 Merate (Lc), Italy

## ABSTRACT

A long-standing problem of astrophysical research is how to simultaneously obtain spectra of thousands of sources randomly positioned in the field of view of a telescope. Digital Micromirror Devices, used as optical switches, provide a most powerful solution allowing to design a new generation of instruments with unprecedented capabilities. We illustrate the key factors (opto-mechanical, cryo-thermal, cosmic radiation environment,...) that constrain the design of DMD-based multi-object spectrographs, with particular emphasis on the IR spectroscopic channel onboard the EUCLID mission, currently considered by the European Space Agency for a 2017 launch date.

## 1. INTRODUCTION

Several of the most compelling astrophysical questions, like the nature of the Dark Energy that accelerates the expansion of the universe or the formation of galaxies, heavily rely on the possibility of taking large samples of spectral data. Eighteenth century astronomers envisioned the classic *objective prism* method, which consists in imaging a celestial field through a prism having the same size of the telescope aperture. This method produces a spectrum of each source in the field. It is especially suited for bright sources, as it presents several drawbacks: 1) the full sky background, integrated over the entire spectrum, falls on each pixel adding to the noise; 2) there is overlap of the spectra of different sources, if they are aligned along the dispersion direction; 3) the effective resolution depends on the apparent size of the celestial object, due to the lack of a slit. Despite these problems, objective prism spectroscopy is still in use because of its simplicity, as it can be added to a conventional imager using a grism, which is a prism with a grating etched on a surface to keep light at a chosen central wavelength undeviated. It can be especially convenient for space applications, due to the low celestial background in comparison with the ground. The imaging instruments onboard the Hubble Space Telescope are typically equipped with one or more grisms. Dedicated satellites performing all-sky surveys in objective prism mode have also been proposed.[1]

In order to eliminate the three problems listed above and reach the ultimate performance, a multi-object spectrograph must be able to operate in slit-mode for each source. Since celestial objects are randomly located, any viable solution must allow for a high degree of versatility. In particular, focal plane masks made by punching small apertures on a metal plate are convenient only if they can be easily exchanged for each celestial field. Since the field projected by a telescope is often affected by geometric distortion, a pre-imaging capability must be present to exactly locate targets in the field. The preparation of the mask requires rather sophisticated (e.g. laser cutting) machining, and the development of mask exchange mechanisms working in cryogenic conditions (for thermal-IR applications) is also complex. Despite these limitations, the removable punch-plate approach has been adopted by a number of instruments.[2,3]

A different strategy uses fibers, positioned on each source with high precision robotic positioners. Using fibers, the spectrograph can be bench-mounted remotely several meters away. The loss of transmission introduced by the fibers and the problems with their positioning in dense fields are largely compensated by the multiplexing advantage. Multi-fiber spectrographs made of several hundreds fibers have been built[4] and this is today the most popular approach for non-cryogenic applications, i.e. in the visible. Other approaches, like the "hybrid" one where the fibers are manually positioned in the holes previously punched in a plate[5] or where a stack of slits is realized by positioning the inner edges of masks sliding from both sides of the field, are also possible.[6,7]

MEMS provide an alternative approach to multi-object spectroscopy. The possibility of having randomly addressable optical switches able to selectively transmit or reflect light to the spectrograph (Figure 1) is extremely attractive. In particular, MEMS allow to envision focal plane masks where the slit pattern can be computer generated in real time, with enormous gains in terms of versatility and convenience.

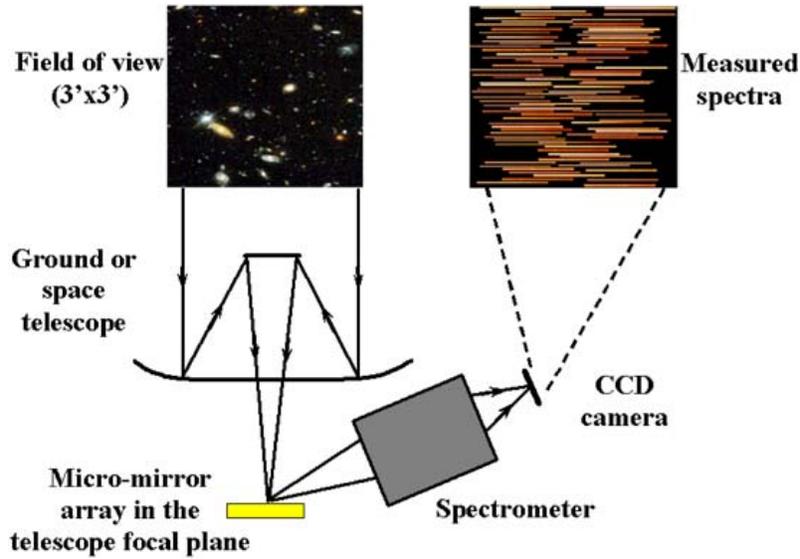

Fig.1 Sketch of multi-object spectrograph based on a micro-mirror (DMD) array. The DMD in the telescope focal plane selects the targets and sends their light to the spectrograph (from http://spie.org/x14024.xml?ArticleID=x14024 ).

## 2. EARLY DEVELOPMENTS

To our knowledge, MEMS devices working either in transmission (Micro-shutter-array, MSA) or reflection (Digital-Micromirror Device, DMD) have been first investigated for astrophysical research by NASA in the late `90s, in the early development phase of the James Webb Space Telescope (JWST, named at that time "Next Generation Space Telescope"). DMDs were first proposed for the JWST multi-object spectrograph.[8] Subsequently, NASA funded both the scientific and technical concept as Pre-Phase A study,[9] together with an exploration of enabling MEMS technology.[10,11] The development of new technology was made necessary by the particular requirements of the JWST instrument (slit size ~100 micron, cryogenic capability, contrast 2000:1 or higher) which made the DMDs developed by Texas Instruments (TI) for commercial projection systems (Figure 2) unsuitable for JWST. At the same time, the NASA Goddard Space Flight Center (GSFC), the Space Telescope Science Institute and the Kitt Peak National Observatory jointly funded the development of IRMOS,[12,13,14] a multi-object near-IR spectrograph based on a TI DMD of 848×600 elements.

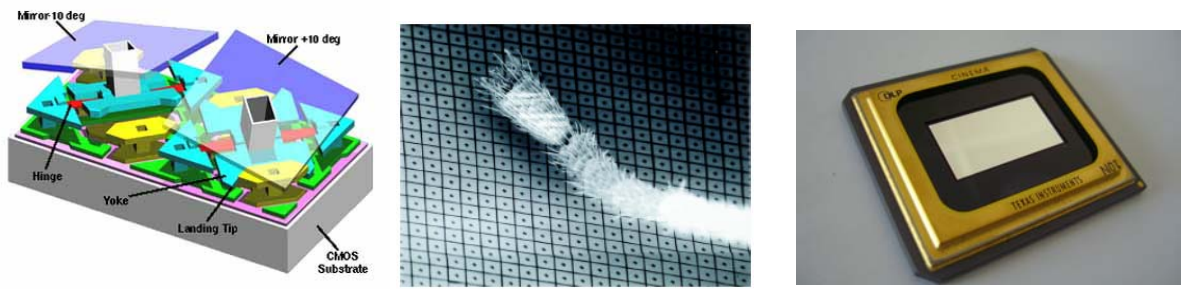

Fig. 2: Left) typical substructure of a TI DMD; center) DMD array with an ant leg for comparison; 3) packaged DMD CINEMA (2048 ×1080) device.

As an alternative to the DMD, NASA/GSFC also developed since 1999 the Micro-Shutter Array. The MSA is an array of small (100×200 micron) shutter blades connected to a frame by narrow torsion bars which can be opened through magnetic actuation and latched open electrostatically.[15,16] The JWST project eventually preferred the MSA to the DMDs, mostly on the basis of easier optical design and concerns relative to scattered light, contrast, and flatness of the relatively large mirror facets. The most recent MSAs qualified for flight on JWST come in 171×365 pixels format and operate in cryogenic conditions (Figure 3).

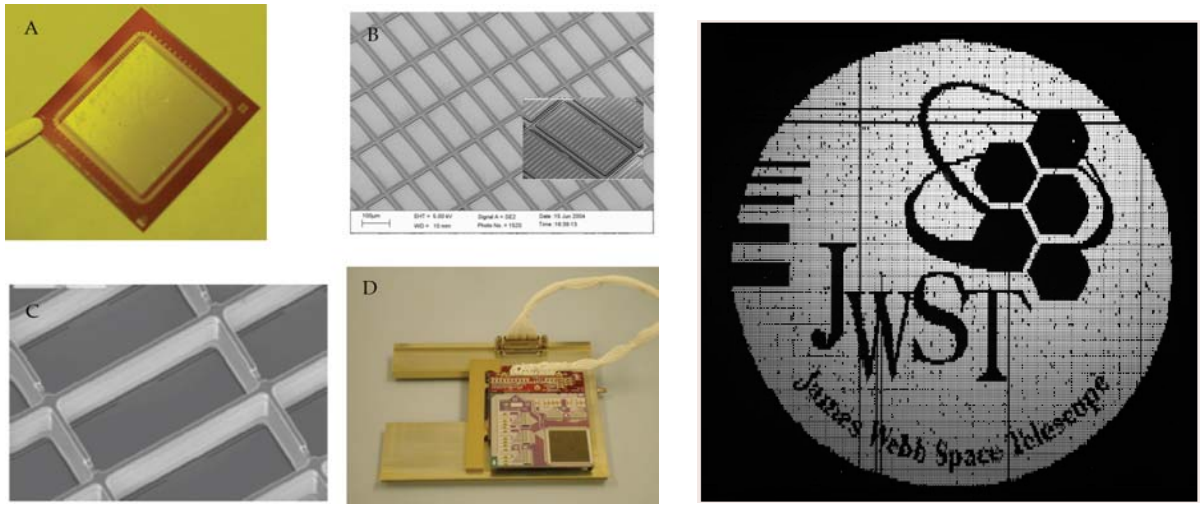

Fig. 3. Left: (A) A microshutter array containing 62,415 individual shutters. (B) SEM image of the front side of a small section of the microshutter array with a SEM image of a single microshutter (inset). (C) SEM image of the backside of the microshutter array. (D) the microshutter array mounted on the silicon substrate with the CMOS ICs (from http://spie.org/x19501.xml?ArticleID=x19501). Right: Micro-shutter array (MSA) programmed with the James Webb Space Telescope symbol (from STScI Newsletter, Summer 2007).

## 3. DMD BASED INSTRUMENTS

The interest of NASA for DMD-based spectrographs echoed in the scientific community, triggering the construction of at least two new astronomical instruments, IRMOS and RITMOS.

### 3.1 IRMOS

The Infra-Red Multi-Object Spectrograph (IRMOS) was conceived within the JWST community to explore the design and performance of a DMD based instrument, while providing at the same time a high-performance scientific instrument for astrophysical research.[12,13,14] IRMOS is based on an early 848x600 element DMD, 10 degrees tilt, each one being a square 16 micron on a side and 17 micron distance between centers. The DMD is cooled at about T=-45C to enable observations from 0.85 to 2.5 microns, thanks to a HgCdTe detector of 1024×1024 pixels. In this wavelength range, there are four main atmospheric windows, respectively z, J, H and K, centered at 0.95, 1.2, 1.6 and 2.2micron. For each window, the spectrograph provides resolution $R=\lambda/\Delta\lambda$ = 300, 1000 and 3000 (1000 only in the z-band), together with some imaging capability. The all-reflective optics of IRMOS deliver a field of view of 170×210 arcseconds at the 4m Mayall telescope of the Kitt Peak National Observatory, where the instrument is currently operated.

A standard IRMOS observation (Figure 4) requires the preliminary acquisition of the targets in imaging mode. Once the positions of the targets have been defined, the micromirrors corresponding to these targets are maintained in their "ON" status while all the others are deflected to "OFF". Then a dispersing element is inserted into the beam and the spectroscopic observation can begin. In practice, an accurate mapping of the DMD-detector transformation is required, which can be easily obtained and monitored by observing a rectangular grid of illuminated micromirrors set to their ON status in imaging mode.

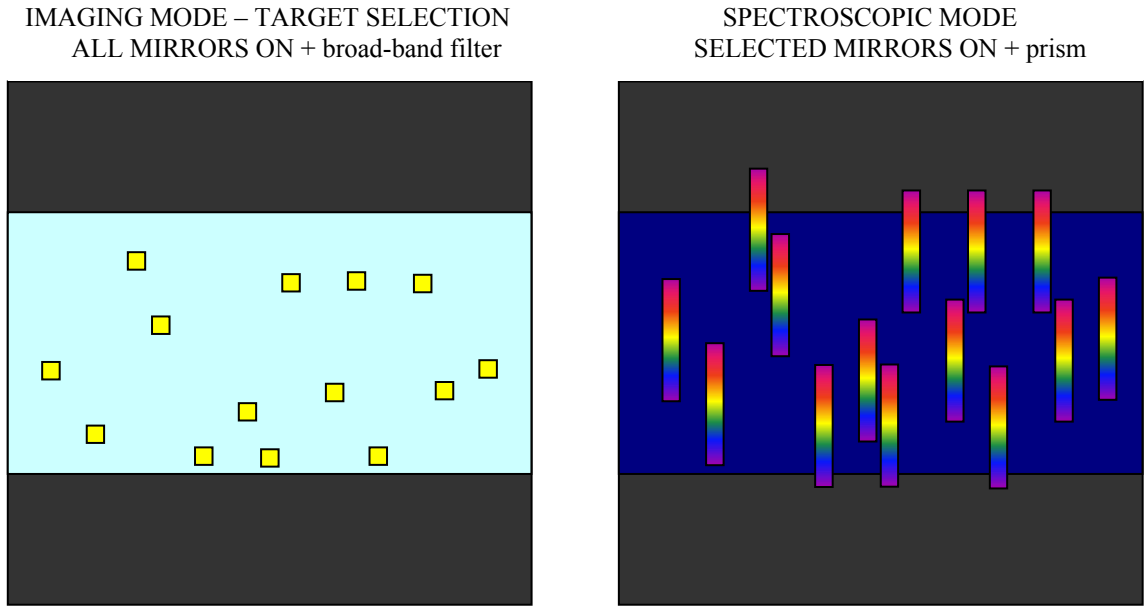

Fig. 4: sketch of the data acquisition and observing procedure. Left: the DMD field is projected onto the detector (dark background) in broad band imaging mode. The light blue color indicates the high background, targets are represented by yellow squares. Right: all DMD mirrors are turned off except those of the targets, the prism is inserted. The background is low (dark blue) and spectra are produced.

The experience with IRMOS has provided a number of insights on the design and operations of DMD based spectrographs. In particular:

1) TI DMDs can be operated at -45C, a temperature well below the nominal operation range of these devices. Low temperature is needed at wavelengths longer than ~1.8 micron to reduce the thermal background introduced by the non reflective parts of the DMD surface, i.e. the gaps between mirrors, which being "black" have high emissivity.

2) The DMD mirrors do not "stick" even after several hours of operations locked in one of the on/off positions. This despite the fact that commercial devices are designed for high frequency (several kilohertz) switching between the two states.

3) The cooling rate of the device must be carefully controlled, the DMD being the warmest component inside the cryogenic instrument. An early failure of a DMD mirror in IRMOS has been attributed to an incorrect cooling curve.

4) Other parameters, like contrast and optical quality, appear to correspond to what is measured by the vendor. In particular, the contrast measured at the telescope of ~300:1 is typical of early generations of DMDs (more recent parts achieve contrast 2500:1, or higher).

IRMOS also allowed envisioning target acquisition and calibration strategies that are unique to MEMS based spectrographs.[17,18] Other operating modes such as Hadamard transforms for integral-field spectroscopy have also been successfully tested.[17] These experiences may be useful for planning the operations of NIRSPEC, the JWST infrared spectrograph based on MSAs. IRMOS has also produced relevant scientific results,[19] matching all original expectations that motivated its development.

## 3.2 RITMOS

The Rochester Institute of Technology Multi-Object-Spectrometer (RITMOS[20]) also utilizes a TI DMD of 848×600 mirrors for multi-object spectroscopy. The main characteristic of RITMOS, which operates at visible wavelengths, is the fact that it exploits both the ON and OFF beams reflected by the DMD. The DMD, located at the telescope focal plane and fed by a telecentric optical system, reflects light to a spectrographic channel on the ON side and to an imaging channel on the OFF side. In the imaging channel, an Offner relay systems reimages the DMD onto a CCD detector. In spectroscopy, a similar relay system feeds a fixed transmission grating which produces visible spectra at about R=6000. In normal operations, the spectroscopic channel is fed by a limited number of mirrors turned ON, whereas the large majority of mirrors, in the OFF status, provide a simultaneous imaging capability useful e.g. to take data on fainter sources below the sensitivity limit of the spectrograph.

RITMOS is a small and light instrument designed for the Mees Observatory 24" Cassegrain telescope of the University of Rochester. Despite the limited sized of the telescope, RITMOS has interesting scientific potential thanks to its clever exploitation of the unique DMD's characteristics.

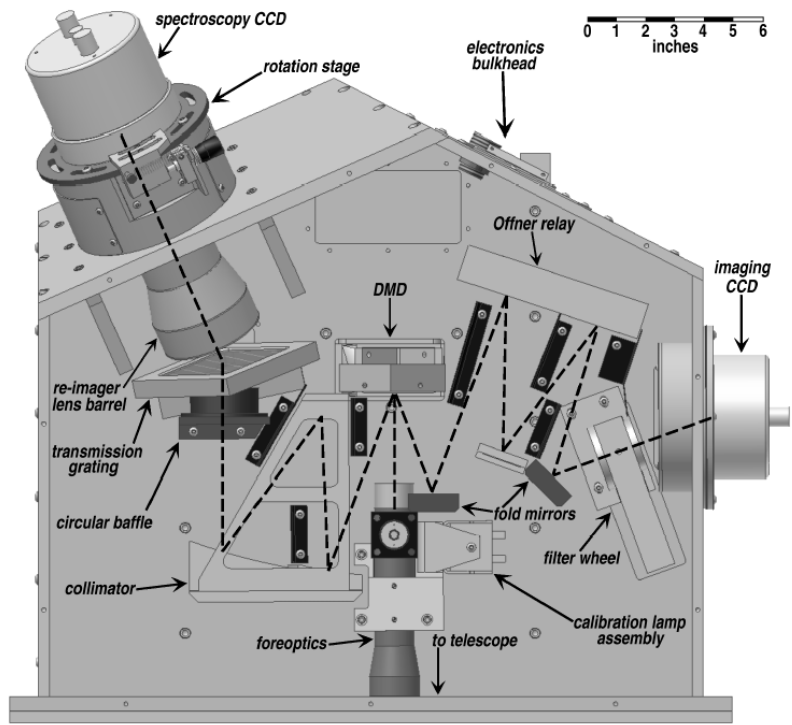

Fig. 5. Layout of the RITMOS spectrograph (from Ref. [20]).

## 4. SPACE/EUCLID

### 4.1 Mission concept

Our team has recently proposed the Spectroscopic All-Sky Cosmic Explorer (SPACE) in response to the Call for Proposal of the European Space Agency for the first planning cycle of the Cosmic Vision 2015-2025 program. The SPACE proposal, envisioned as a joint ESA/NASA collaboration, has passed the first ESA selection. The original concept has evolved into the Near-Infrared Spectrograph channel of the EUCLID satellite, under consideration by ESA for a 2017 launch date. As the definition of the EUCLID Near-Infrared Spectroscopic (ENIS) channel has not yet been completed, hereafter we will mostly refer to the original SPACE proposal in what concerns the technical aspects, whereas for the science performance we will refer to the most recent EUCLID/ENIS estimates.

The main scientific goal motivating SPACE/ENIS[21,22] is to produce the largest three-dimensional evolutionary map of the Universe over the past 10 billion years. In the current configuration, it is expected that ENIS will take near-IR spectra of about $2\times10^9$ galaxies over the half of the sky unobstructed by the Galaxy. ENIS will precisely locate ($\Delta z\sim0.001$) each galaxy and observe baryonic acoustic oscillation (BAO) patterns in the Universe between 5 to 10 billion years ago. The BAO is a scale in the spatial power spectra which can be used as a "standard ruler". The BAO evolution across the history of time allows measuring the equation of state and rate of change of dark energy. Reconstructing the 3-d structure of the Universe will also allow an accurate assessment of the evolution of structure formation over the last 10 billion years, providing a complementary method to discriminate between theories of dark energy and theories of modified gravity invented to explain the acceleration of the universe.

To perform the deepest "all sky" imaging survey ever attempted in the near-IR, SPACE/ENIS will rely on DMDs. The ESA requirement of TRL-6 technological maturity imposes the selection of TI devices already available. We have opted for the largest possible devices, i.e. the TI Cinema chip of 2048×1080 elements, 13.6micron on a side, with 12 degrees tilt angle around the diagonal. For the SPACE proposal we designed an optical system based on a 1.5m telescope primary mirror, with fore-optics delivering 0.75arcseconds/pixel intermediate field of view on the micromirror, sampled with 2x2 pixels of a HgCdTe detector with sensitivity in the near-IR up to 1.7micron. In order to cover the largest possible field per telescope pointing, we envisioned an array of four identical spectrographs (reduced to 3 in EUCLID/NIS, Figure 6), each one fed by a face of pyramid folding mirror placed in the vicinity of the telescope focal plane. This arrangement, also used in previous HST instruments, provides contiguous spatial coverage when detectors (or, in our case, DMDs) cannot be easily butted together.

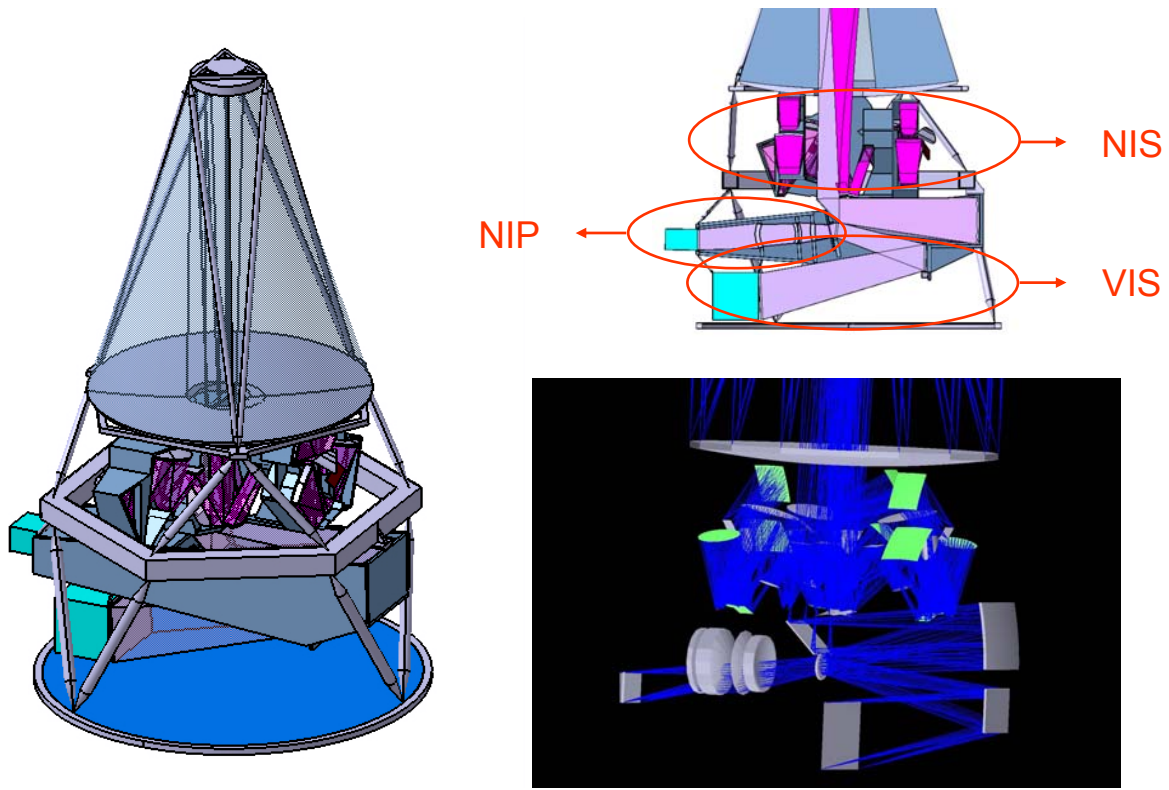

Fig. 6. Left: conceptual design of EUCLID spacecraft, based on a 1.2m 3 mirror Korsh reflector. Top-right: side view of the preliminary optical design. NIS (Near-Infrared Spectrograph) is the DMD-based instrument discussed in this paper. NIP and VIS indicate the wide field imaging channels for Near-IR photometry and Visible imaging, respectively. Bottom-right: 3-d view of the instrument setup, with the 3 NIS spectrographs on top.

### 4.2 DMD space qualification

To minimize the IR background and operate in a most stable thermal environment, EUCLID will fly at the 2nd Lagrangian point of the Earth-Sun orbit, at approximately 1.5 million kilometers of distance from the Earth. At this location, the payload will be fully exposed to the flux of high energy solar particles, without the beneficial shielding of the Van Allen belts encountered by e.g. the HST. Therefore, a critical aspect of the SPACE/ENIS mission is the space qualification of commercial DMDs, as well as of their control electronics (all under the common denomination of DC2K DLP Chip Set). A consortium has been organized between Visitech (Norway) and Laboratoire d'Astrophysique de Marseille (LAM). Visitech, a world wide TI distributor and a leading Digital Light Processing (DLP®) product design and development company, will provide the printed circuit boards holding the Chip Set and all supporting electronics designed specifically for these tests, besides providing managerial overview. LAM, one of the major astrophysical laboratories in France, participated in early tests of DMDs for JWST/NIRSPEC commissioned by ESA gaining a solid expertise in MEMS development and testing for the next generation of astronomical instrumentation. LAM's main role is to design and realize the cryostat and optical bench used to perform environmental testing (ambient and cryo-vacuum) and complete characterization of the DMDs before, during and after the different tests. Part of the tests will also be executed at IASF-Bologna (Italy) (vibration) and ESA (radiation). The preparation of the tests is under way and results are expected in the Fall of 2009.

In general, one would expect MEMS devices to be radiation hard by default. Even at the high end of space mission doses, the mechanical properties of silicon and metals are mostly unchanged (Young's modulus and yield strength not significantly affected). Dielectric charging remains the dominant failure mechanism in electrostatically operated MEMS. An analysis of the published radiation tests on MEMS devices (either single crystal or poly-silicon) shows that the minimum dose for failure ranges between 20 kRad and several MRad.[23] The lowest value, 20 kRad, can be compared with the 2kRad of ionizing radiation per year near solar maximum (worse case scenario) estimated for the shielded SNAP CCD detectors. It is worth to notice that EUCLID will fly at the minimum of the solar cycle. In principle, some radiation damage from heavy ions may cause catastrophic latch-up in CMOS, but this could be readily prevented using a latch-up protection circuit that senses the increase in power draw and shuts down the device. In any case, it is the CMOS underlying the DMD mirrors which will most probably drive the requirement on radiation shielding.

### 4.3 Other DMD issues for space application

The susceptibility of DMDs to cosmic radiation is probably the most critical parameter for space qualification. In what concerns the other parameters, DMDs appear extremely reliable due to their technological maturity. In particular:

1) DMDs normally show a lifetime of over 100,000h with no degradation, with trillions of cycles perfomed by each element. Typical astronomical applications require locking the mirrors for several minutes, at least, to allow for signal integration. This operating mode eliminates all concerns related to nanoscale fatigue and fracture of the mechanisms.

2) While they were an issue in early generations of DMDs, stuck pixels have now virtually disappeared. Since typical projection applications do not forgive dead pixels, DMDs typically show ~1 dead pixel per million, a yield higher than that achieved by state-of-the-art 2-d science detectors. Moreover, dead pixels remain in the flat (zero position) state, i.e. they do not represent a source of unwanted leaking light.

3) The DMD elements, designed to oscillate at ~10kHz, have mechanical parameters (e.g. resonant modes) that make them able to resist extremely high levels of vibration and acceleration. The macroscopic DMD enclosure, rather than the microscopic DMD structure, may actually represent the major cause of concern during launch.

4) In high-end 3 chip DLP projectors, 5-10kW Xe discharge lamps are mounted in the immediate vicinity of the DMDs, which therefore operate without problems in rather harsh thermal environments (up to +80C) and under very high radiation fluxes, from the near-UV to the IR. At the other extreme, we know that the old DMD installed on IRMOS is routinely operated and thermally cycled at -40-50C in cryo-vacuum conditions.

Another set of critical points concerns the optical performance of DMDs. To fully exploit the unique advantages of the extraterrestrial environment, space instrumentation must preserve the extremely low sky background and diffraction limited optical quality allowed by the absence of the turbulent atmospheric medium. Having an intermediate focal plane populated with micromirror (or microshutter) facets may potentially degrade the overall sensitivity of the instrument in a number of ways:

5) Diffraction: DMDs facets receive a nearly diffraction-limited point spread function (PSF) from the foreoptics. If the PSF falls on adjacent mirrors, some diffraction will be added due to the grating effect of the mirror edges, plus the double phase delay caused by the mirror tilt. The net result is a broadening of the light distribution at the spectrograph pupil, which due to its finite size will transmit only a fraction of the beam. Also, the location of a point source in the slit has a major impact on the fraction of light transmitted to the spectrograph. To minimize these effects, it is beneficial to have micromirrors larger than the PSF ($\lambda/D=0.2''$ at $\lambda= 1$ micron for a 1m aperture telescope) as well as an oversize of the pupil, dispersing element and part of the camera optics. In practice, the light losses will depend on the position of the source on the DMD and finding the optimal compromise requires detailed simulation. The problem seems to be worse for spectrographs operating in the thermal-IR regime, as the requirement of a tight cold stop at the exit pupil conflicts with the need of accepting a larger beam to reduce diffraction losses. This is not the situation in which SPACE/ENIS will operate.

6) Emissivity: the DMD filling factor, approximately 90%, suggests a ~10% emissivity (the actual value depends on the actual tilt angle of the device in the optical beam). There is therefore a thermal emission which strongly depends on the DMD operating temperature, rising exponentially with the wavelength. Trade-off studies are needed to optimize the DMD temperature and the long wavelength cutoff of the detector and filters.

7) Image quality: besides diffraction, the DMD has very little effect on the image quality. The tilt can create a small defocus, but this is a higher order effect. The DMD tilt angle is limited by a mechanical stop, providing excellent uniformity and consistent image quality for the entire lifetime of the device. Due to their small size, the individual mirror facets should have extremely low distortion.

8) Contrast: major improvements have been made in recent years, and commercial DLP systems show today contrast ratios higher than 2500:1, which in astronomical terms translates in 8.5 magnitudes of rejection. This value may depend on the f/# of the illuminating beam, but the fast values typically adopted in DLP projectors appear well matched to the astronomical requirements that normally push for large field of views, i.e. fast f/#.

9) Coating: TI offers for the standard 1024×768 DMD parts three different coatings optimized for UV, optical, and IR applications. The flight parts for SPACE/ENIS will be IR coated.

In summary, DMDs appear both mechanically and thermally reliable. Radiation effects, even in the worst case scenario, seem to be manageable with standard techniques. If this will be confirmed by the tests to be performed in Europe for the ENIS program, TI DMDs will definitely be regarded as a most viable tool for a new generation of space instrumentation.

## 5. FUTURE PROSPECTS

In this final section we report on a few recent developments in the field of DMDs that, enabling new instrument concepts, may open new possibilities of research.

### 5.1 Concept of new instruments

The fact that DMDs allow for new observing capabilities is nicely demonstrated by the RITMOS design (Section 3.2), which allows for parallel imaging and multi-object spectroscopy of the same field. Several optical solutions have been explored for the SPACE spectrograph,[24] finding ways of reducing the number of mirrors, adding an imaging capability, eliminating the fore-optics, and exploring the use of Total Internal Reflection (TIR) prisms to substantially reduce the packaging. A new type of multi-object spectrograph with OH suppression capability based on two DMDs has also been envisioned:[25] the first DMD is used to select the sources, whereas the second one receives a highly dispersed spectrum and selectively reflects only those wavelengths not affected by atmospheric OH airglow into a lower resolution camera/detector system, for increased signal to noise. The range of observing capabilities opened by MEMS devices, and DMDs in particular, remains to be fully exploited.

## 5.2 Concept of new observing modes

In Section 3.1 we have already mentioned the possibility of operating the DMDs as Hadamard coded masks. The validity of this approach has been demonstrated with IRMOS, but other groups have also considered this technique[26] for applications other than astronomy. In Hadamard spectroscopy, several slits are opened at the same time collecting overlapped spectra. By properly selecting the slit pattern, it is possible to obtain a set of combined spectra which can be readily inverted, providing for each pixel, and therefore for the full field, the entire spectral information. A Hadamard transform spectrograph allows observing in integral field mode areas much larger that those typically mapped with image slicers. To be effective, this method requires high stability and low readout noise; both conditions can be easily attained in space, especially if one can adopt relatively low spectral resolution to keep the readout noise negligible. The original SPACE proposal included a unique galactic plane survey, covering in Integral Field (Hadamard) mode the strip ±0.5° centered on the Galactic Equator between ±60° of Galactic Longitude.

## 5.3 Development of new DMDs for science use

We are aware of at least three programs to independently develop micromirror devices for space science. In the US, both NASA/Goddard and Sandia National Laboratories participated in the past in the early development phase for NGST/JWST.[11,27] In Europe, an effort is currently under way to develop single-crystalline silicon micromirror arrays for future generation infrared multiobject spectroscopy.[28,29,30] These are micromirrors of 100μm × 200μm in size which can be tilted by electrostatic actuation by 20°. Gold-coated arrays of 5x5 micromirrors have been tested below 100K. Both coated and uncoated micromirrors are optically flat (peak-to-valley deformation $<\lambda/20$ for $\lambda>1$micron) at room temperature and in cryogenic environment. Successful actuation has been done at room temperature and at temperatures below 100K. Large arrays of 200x100 micromirrors are being fabricated and an actuation scheme for extremely large arrays has been developed. In general, this program seems able to deliver in the future parts matching all the initial requirements of JWST/IRMOS.

## 6. CONCLUSIONS

In this paper we have shown how DMDs provide an ideal solution to the longstanding quest of astronomy for simultaneously obtaining spectra of thousands of faint sources randomly positioned in the field of view of a telescope. DMDs have been considered for the next NASA flagship mission, JWST, and currently by ESA for the EUCLID satellite aimed to map the 3-d structure of the universe. We have discussed the key factors that constrain the design of DMD-based multi-object spectrographs, with special regard to space applications. We have also illustrated the possible future developments enabled by this remarkable technology.